\documentclass{kluwer}
\begin{document}
\begin{article}
\begin{opening}

\title{FORMATION OF MASSIVE COUNTER-ROTATING DISKS: AN ALTERNATIVE SCENARIO}

\author{Iv\^anio \surname{Puerari}\email{puerari@inaoep.mx}}
\institute{Instituto Nacional de Astrof\'\i sica, Optica y Electr\'onica,
Calle Luis Enrique Erro No. 1, 72840, Tonantzintla, Mexico}

\author{Daniel \surname{Pfenniger}\email{daniel.pfenniger@obs.unige.ch}}
\institute{Observatoire de Gen\`eve, Universit\'e de Gen\`eve,
CH-1290 Sauverny, SWITZERLAND}

\end{opening}

\section{Introduction}

The formation of massive counter-rotating disks
is not yet fully understood. It is unlikely that counter-rotating
systems result from the galaxy formation process, because in a uniformly
rotating protogalactic cloud a subsequent splitting of the angular
momentum distribution into a strongly bimodal one appears
``anti-thermodynamical''.
More 
natural is to look for a second event scenario in the form of a merger
process as the origin of the
formation of this kind of systems, because infalling satellites do
contain much specific angular momentum with arbitrary orientation.
Although only a few cases
of {\sl massive} counter-rotating disks are known,
counter-rotation in galaxies seems
to be a common phenomenon, particulary in early-type spirals (Kuijken,
Fisher \& Merrifield 1996; for reviews, see Rubin 1994, and Galletta
1996).
So systems with heavy counter-rotation might represent the extreme case of
the lighter but more frequent counter-rotating systems.

Thakar \& Ryder (1996, 1998) tested some scenarios for the formation
of counter-rotating disks. They investigated three mechanisms: episodic
gas infall, continuous gas infall, and merger with a gas-rich dwarf
companion. In all cases, some counter-rotation was produced, but in some
cases they had very different characteristics when compared to galaxies
like \hbox{NGC 4550}, the first galaxy with massive counter-rotating
disks discovered (Rubin, Graham \& Kenney 1992). For example,
with gas infall, Thakar \& Ryder were able to produce small counter-rotating
disks with radial profile different from exponential. Yet, some conditions
must be satisfied for producing counter-rotation, without upsetting
the stability of the existing disk: e.g., the rate of infall has to be
small to preclude excessive heating. The scenario with gas-rich dwarf
merger is even less viable to be a mechanism to produce 
massive counter-rotating
disks: only very small dwarf galaxies do not increase significantly 
the thickness of the
pre-existing disk, and the timescale for this process is prohibitively
long.  
In addition, merging several dwarf galaxies over several Gyr requires
to adjust their angular momentum vectors relative to the moving large
galaxy such that only when they merge their angular momentum vectors
are well correlated, which appears highly unprobable.

In this contribution we will investigate an alternative scenario:
{\sl major} mergers between progenitors with comparable mass. 
Such a scenario can be ruled out in general because they produce 
rather ellipticals.  Yet for a 
narrow range of initial conditions, {\sl major}
mergers can nevertheless produce anyway rare galaxies 
such as \hbox{NGC 4550}, and so
they are viable alternatives for the formation
of these counter-rotating systems. The scenario
is successful in producing a remarkably axisymmetric disk possessing
strongly counter-rotating populations. Partial results were already published
by Pfenniger (1997). Here, we will present results for different
galaxy models and for simulations including gas.

\section{Models and computational details}

Fully self-consistent models with dark halo were constructed using a
recipe similar to that of Barnes (1988). Firstly, we construct the spherical
system following a Plummer's density law, secondly, the disk,
following
a radial Kuzmin/Toomre's law, and a $\mathop{\rm sech}^2(z/z_0)$ law in the
vertical direction.
The next step is to tabulate the forces from the disk distribution on a grid.
This force grid is slowly imposed to the spherical system. When the
spheroid reaches
equilibrium, we calculate the rotation curve from this evolved halo, and
give velocities to the disk particles, chosing velocity dispersions
($Q$ Toomre's parameter equals 1, and epicyclic aproximation). The asymmetric
drift correction is taken into account. By superposing this disk and
the evolved halo, we get an isolated galaxy model, close to equilibrium.
In Table I we give the parameters for the isolated model.
$M_H$ and $M_D$ represent the halo and disk mass, respectively.
$N$ is the number of particles, and $b$ the radial scale length.
$R_{cut}$ is the cut-off radius, and finally $z_0$ is the vertical
scale height for the disk. With the chosen normalization ($G=1$),
the units of length, time and mass are 3 kpc, $10^7$ years,
and $6\times 10^{10}\ {\rm M}\odot$.

\begin{table}
\caption{Model parameters}
\begin{tabular}{lllllllll}\hline
Halo &&&& Disk &&&&\\ 
\lcline{1-4}\rcline{5-9}
$N_H$ & $M_H$ & $b_H$ & $R_{cut_H}$ & 
$N_D$ & $M_D$ & $b_D$ & $R_{{cut}_{D}}$ & $z_0$ \\
30000 & 1.5 & 5.0 & 10.0 & 10000 & 0.5 & 1.0 & 5.0 & 0.1\\
\hline
\end{tabular}
\end{table}

\vskip20pt
The binary galaxy systems are set by rotating and displacing the isolated
model to get the desired configuration. We choose coplanar,
antiparallel systems, since this configuration minimize the energy which would
contribute to the heating of the disks.
To check the effect of the orbit, we have run a
parabolic and a circular encounter. The initial distance between galaxies is
22 lenght units, and for the parabolic orbit, the pericenter distance
is 6 length units (the disk initial radius is 5). These simulations
have been calculated using a parallellized version of the {\sl treecode}
(Barnes \& Hut 1986), running on a Power Challenge Silicon Graphics.
We have run also two simulations including gas. On these runs,
10\% of one disk mass is treated as gas in a {\sl tree-sph}
code. The orbit for these simulations is parabolic.   In one case, the
gas is in the prograde galaxy, and in the retrograde galaxy in the
second simulation. More details of the run parameters will be given
elsewhere (Pfenniger \& Puerari, in preparation).

\section{Results and discussion}

We have plotted the average line profiles
as seen in the edge-on disks. The bimodal distribution is crucial
in order to distinguish counter-rotating disks from hot population
with a zero net rotation. It is clear from our velocity
profiles (not shown here) that the distribution is bimodal,
representing counter-rotating systems very well.

In Figure 1 we plot the tangential velocity
(as seen in the edge-on disks)
of the disk particles, in the case of a parabolic orbit. 
This plot can be compared to the rotation
curves derived for \hbox{NGC 4550} (Rubin et al. 1992) and
\hbox{NGC 3593} (Bertola et al. 1996). The similarity
between the computed and the measured rotation curves
show that a {\sl major merger} scenario can reproduce very
well the kinematical characteristics of the counter-rotating galaxies.

Figure 2 is similar to Fig. 1, but now for a circular orbit.
The ratio between the ``prograde'' and ``retrograde'' rotation velocities
seems to be related with the orbit: a more circular one
tends to yield low counter-rotating velocities, while
parabolic orbits yield high counter-rotation.

\begin{figure}
\includegraphics{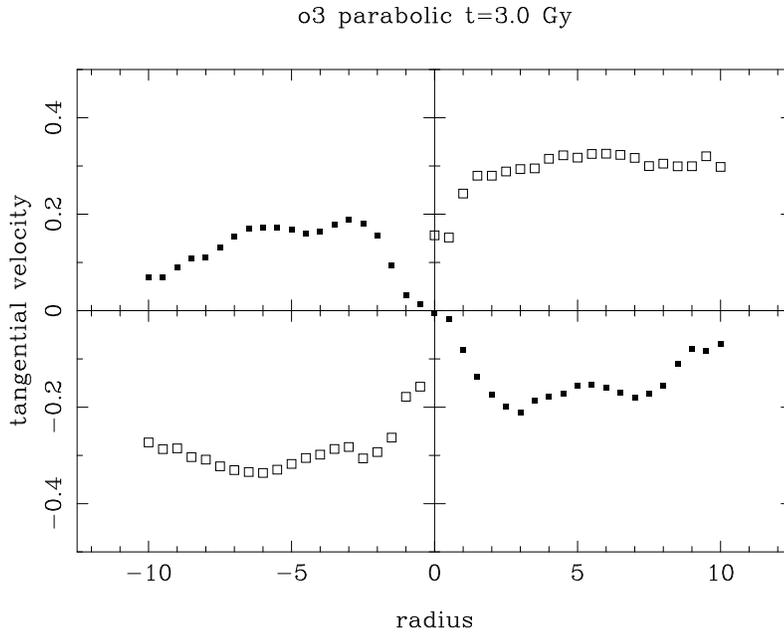}
\vskip8truecm
\caption{Rotation velocity curve of the disk particles of the merger
remnant for the case of parabolic orbit to be compared with the velocity
curve of \hbox{NGC 4550} (see, e.g., Fig. 3 on Rubin et al. 1992) or
\hbox{NGC 3593} (see, e.g., Fig. 1 on Bertola et al. 1996).}
\end{figure}

With respect to the density profile, our models are constructed following
initially a Kuzmin/Toomre density law. This profile remains
almost unchanged after the merger. So, our models have
no difficulty to reproduce normal radial density profiles
after the merger.

\begin{figure}
\includegraphics{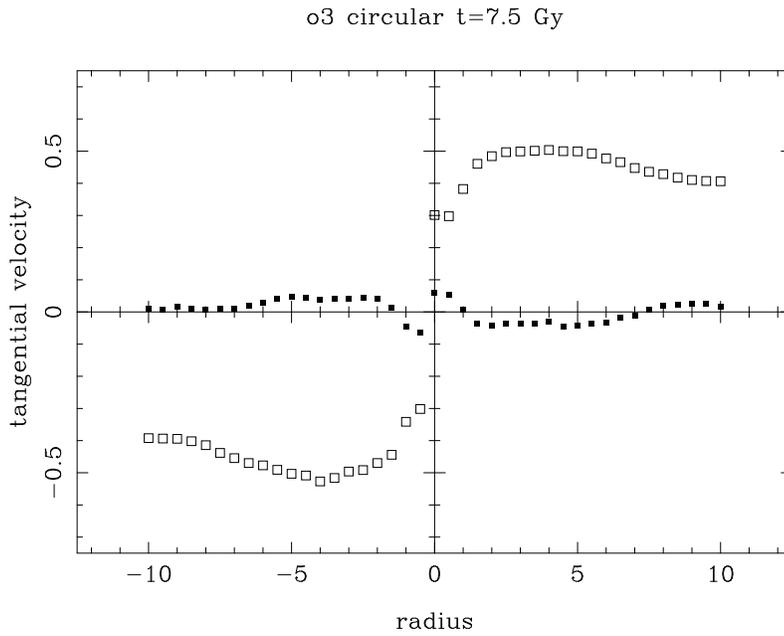}
\vskip8truecm
\caption{The same as Figure 1, but now for a circular orbit. 
The amplitude of the counter-rotation is less marked compared
to the parabolic case.}
\end{figure}

At some stage of the evolution and for some edge-on viewing angles,
coplanar disks can appear as edge-on galaxies with
``two-bulges''. Such a bizarre case is known in the Hercules cluster,
\hbox{PGC 57064}. Incidentally, close-by is a pair of contact spirals
(NGC 6050), which triggered for a part this work.  
For a comparison, we plot in Figure 3,
the image of PGC 57064, taken from the Digitized Sky Survey, and a
snapshot of one of our simulations.  Candidates for
merging coplanar-antiparallel disks such as \hbox{PGC 57064}
could be investigated by spectroscopy or adaptive optics photometry
with large telescopes.

\begin{figure}
\includegraphics{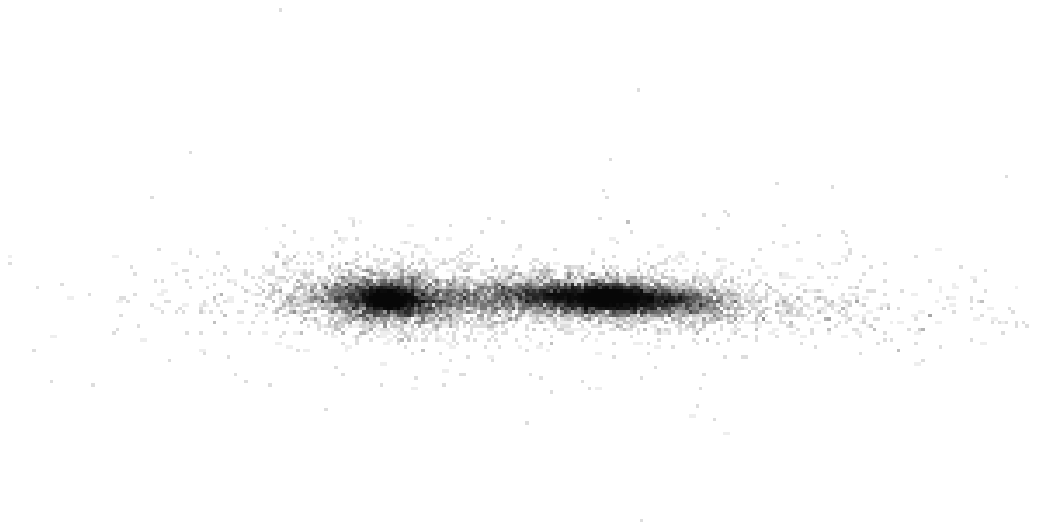}
\vskip7truecm
\includegraphics{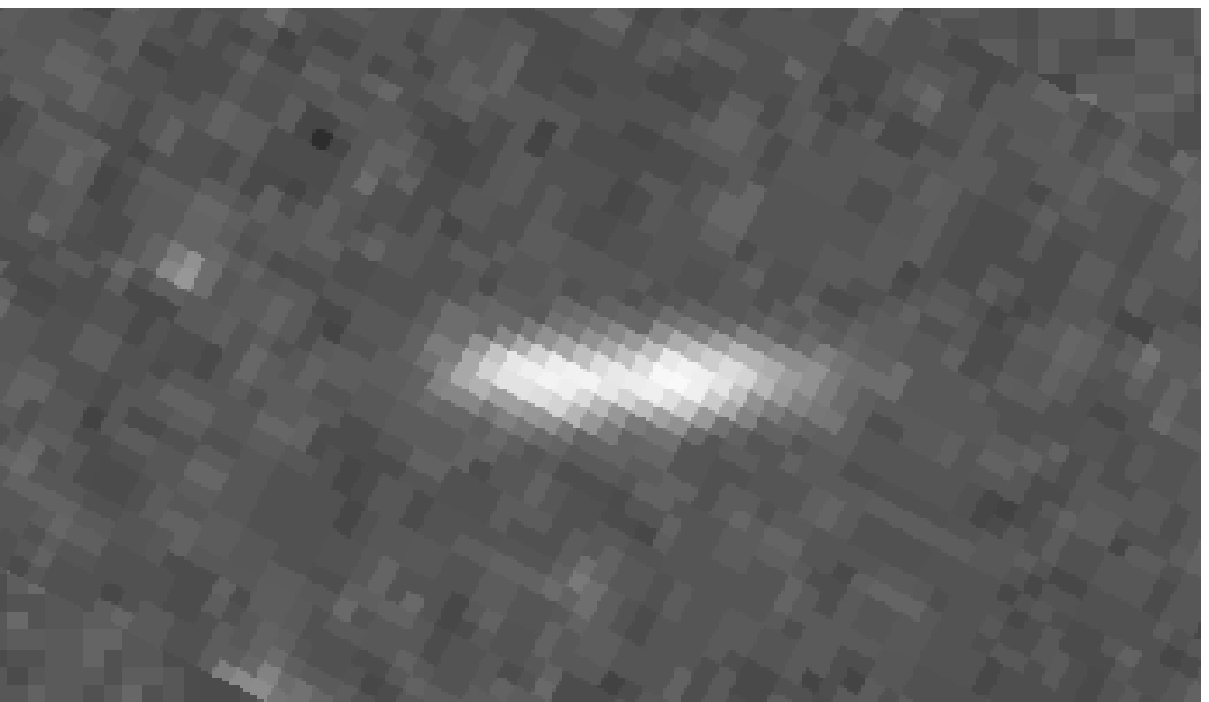}
\vskip7truecm
\caption{Top: Edge-on view of a simulation at some stage of the interaction,
before the merger. This particular view can represent ``two-bulges'' 
galaxies like \hbox{PGC 57064} (botton).}
\end{figure}

It is clear that the retrograde disk galaxy (with respect to the 
orbital spin) is less affected by the interaction: the large tail appears
only in the prograde galaxy. So, the final content of gas in such
interactions depends on where the gas was before the merger.
In typical spirals much gas exists in the outer disks so a coplanar
merger is likely to eject much of it, evacuating simultaneously
the excess angular momentum.

\section{Conclusions}

We have shown that {\sl major} mergers, involving disk galaxies
with comparable masses, can reproduce very well the morphological and
the kinematical characteristics of {\sl massive} counter-rotating
galaxies
such as NGC 4550, a prototype of this kind of objets.
The rotation curve, velocity dispersion and density profile
of the simulations mimic those of counter-rotating systems,
and these counter-rotating characteristcs remain for some
Gyr. The somewhat peculiar conditions for the encounters are coplanar,
antiparallel disks, that should be favored by gaseous viscous processes
taking place when the outer disks interact.  We have tested both 
cases with or witout hot dark halos: they don't appear to be necessary
in the scenario.

The gas content on the final merger depends on where the gas
is before the interaction. If the gas is on the prograde
galaxy, a large amount is expelled, while more
gas remains in the merger remnant if the gas is initially in
the retrograde galaxy.

Candidates where the massive counter-rotating disks
are at earlier stages
of the merger could be galaxies like \hbox{PGC 57064}
(these systems seem to be edge-on galaxies with ``two-bulges'').
Spectroscopy with large telescopes could check the antiparallelism
of such systems and provide some hint about the frequency of such
mergers events.


\end{article}
\end{document}